\documentclass[twocolumn,aps,prl,showpacs,floatfix]{revtex4}

\usepackage{graphicx}% Include figure files
\usepackage{dcolumn}% Align table columns on decimal point
\usepackage{bm}% bold math
\usepackage{latexsym}

\begin{document}

\title{Determination of the strange form factors of the nucleon \\
from $\nu p$, $\bar{\nu} p$, and parity-violating $\vec{e}p$ elastic scattering}

\author{Stephen F. Pate}
 \email{pate@nmsu.edu}
\affiliation{Physics Department, New Mexico State University, Las Cruces NM 88003}

\date{\today}

\begin{abstract}
  A new method of obtaining the strange form factors of the nucleon is
  presented, in which forward-angle parity-violating $\vec{e}p$
  elastic scattering data is combined with $\nu p$ and $\bar{\nu} p$
  elastic scattering data.  The axial form factor in electron-nucleon
  scattering is complicated by the presence of electro-weak radiative
  corrections that in principle need to be calculated or separately
  measured, but this axial form factor is suppressed at forward
  angles.  The neutrino data has no such complication.  Hence the use
  of forward-angle parity-violating $\vec{e}p$ data with $\nu p$ and
  $\bar{\nu} p$ data allows the extraction of all three strange form
  factors: electric, magnetic and axial ($G_E^s$, $G_M^s$, and
  $G_A^s$).  In this letter, $\nu p$ and $\bar{\nu} p$ data from the
  Brookhaven E734 experiment are combined with the Jefferson Lab
  HAPPEX $\vec{e}p$ data to obtain two distinct solutions for the
  strange form factors at $Q^2$ = 0.5 GeV$^2$.  More generally,
  combining the neutrino elastic scattering data from E734 with the
  existing and upcoming $\vec{e}p$ data will yield the strange form
  factors of the nucleon for $Q^2$ of 0.45-1.05 GeV$^2$.  Measurement
  of $G_A^s$ is crucial to the determination of the strange quark
  contribution to the nucleon spin, $\Delta s$.
\end{abstract}

\maketitle

In the last 15 years, a program of parity-violating polarized electron
elastic scattering experiments has evolved, with the goal of measuring
the strange form factors of the nucleon.  These experiments build on
the numerous existing (and ongoing) measurements of the neutron and
proton vector electromagnetic form factors, $G_E^n$, $G_M^n$, $G_E^p$,
and $G_M^p$.  They gain sensitivity to the strangeness contribution to
these form factors by exploiting the interference between
electromagnetic and weak processes.  This interference produces a
helicity-dependent parity-violating asymmetry in the elastic
scattering.  This program of measurements was sparked by the seminal
papers of Kaplan and Manohar~\cite{Kaplan:1988ku},
McKeown~\cite{Mckeown:1989ir}, and Beck~\cite{Beck:1989tg}.  The
SAMPLE experiment at the MIT-Bates Lab and the HAPPEX experiment at
Jefferson Lab have already made measurements of these asymmetries, and
a tremendous amount of additional data will be collected on these
asymmetries at Mainz~\cite{PVA4} and Jefferson
Lab~\cite{G0:Beck,E99115,E00114,E91004} in the next 3-4 years.

This program of experiments has been designed to measure the strange
{\em electric} and {\em magnetic} form factors.  It is desirable to
measure also the strange {\em axial} form factor, $G_A^s$, but there
is a complication.  The parity-violating asymmetry observed in these
experiments, when the target is a proton, can be expressed as
\cite{Musolf:1994tb}
\begin{eqnarray*}
A &=&
\left[\frac{-G_F Q^2}{4\pi\alpha\sqrt{2}}\right] \\
&\times&
\frac{\epsilon G_E^\gamma G_E^Z + \tau G_M^\gamma G_M^Z - (1-4\sin^2\theta_W)\epsilon' G_M^\gamma G_A^e}
     {\epsilon\left(G_E^\gamma\right)^2 + \tau\left(G_M^\gamma\right)^2}
\end{eqnarray*}
where $G_{E(M)}^\gamma$ are the traditional electric (magnetic) form
factors of the proton and $G_{E(M)}^Z$ are their weak (Z-exchange)
analogs, $\tau=Q^2/4M_p^2$, $M_p$ is the mass of the proton,
$\epsilon=[1+2(1+\tau)\tan^2(\theta/2)]^{-1}$, $\theta$ is the
electron scattering angle, and
$\epsilon'=\sqrt{\tau(1+\tau)(1-\epsilon^2)}$.  Lastly, $G_A^e$ is the
effective axial form factor seen in electron scattering:
$$ G_A^e = -G_A^{CC} + G_A^s + G_A^{\rm ewm}. $$ 
Here, $G_A^{CC}$ is the charged-current axial form factor, which has
been measured in neutrino scattering and electroproduction of
pions~\cite{Liesenfeld:1999mv}: at $Q^2=0$ it equals the axial-vector
coupling constant, $g_{A} = 1.267 \pm 0.0035$, and its
$Q^2$-dependence is well described by a dipole form,
$g_{A}/(1+Q^2/M_A^2)^2$, where $M_A = 1.026 \pm 0.021$ GeV is the
axial mass.  The term $G_A^s$ is the strange axial form factor: at
$Q^2=0$ it equals $\Delta s$, the first moment of the spin-dependent
strange quark momentum distribution measured in polarized
deep-inelastic scattering, from which data is it
estimated~\cite{Filippone:2001ux} that $\Delta s = -0.14 \pm 0.03$.
The $Q^2$-dependence of $G_A^s$ is unknown.  Lastly, $G_A^{\rm ewm}$
represents electroweak mixing contributions to $G_A^e$; this includes
radiative corrections to Z-exchange diagrams.  This last term
$G_A^{\rm ewm}$ must be calculated in order to allow extraction of
$G_A^s$ from a measurement of $G_A^e$ in elastic electron scattering.
The difficulties in calculating this term have been noted
already~\cite{Musolf:1990ts,Musolf:1992xm,Musolf:1994tb,Zhu:2000gn},
and have played a role in recent experimental work.  The SAMPLE
experiment, which combined backward-angle electron-scattering data on
proton and deuteron targets to extract $G_M^s$ and $G_A^e$ at
$Q^2=0.091$ GeV$^2$, originally reported~\cite{Hasty:2001ep} values of
$G_M^s = 0.14 \pm 0.29 \pm 0.31$ and $G_A^e = 0.22 \pm 0.45 \pm 0.39$,
which deviated from the calculated value~\cite{Zhu:2000gn} of $G_A^e$
by $1.5\sigma$ and had the opposite sign.  A new analysis of the
SAMPLE data at $Q^2=0.091$ GeV$^2$, along with a new measurement at
$Q^2=0.038$ GeV$^2$~\cite{Ito:2003mr}, now supports the theoretical
results of Zhu et. al~\cite{Zhu:2000gn} at these very low values of
$Q^2$, while not significantly changing the value of $G_M^s$ at
$Q^2$=0.091 GeV$^2$.  However, it is important to note that the value
of $\Delta s$ used in the interpretation of these data is still only
an estimate from polarized deep-inelastic scattering experiments.
Better experimental information on the strange axial form-factor and
$\Delta s$ is needed for a final determination of the strange magnetic
form factor from these data.

The $G^0$ Experiment~\cite{G0:Beck} at Jefferson Lab will circumvent
this difficulty with the axial term by combining three measurements:
forward scattering of protons from $\vec{e}p$ collisions, backward
scattering of electrons from $\vec{e}p$ collisions, and backward
scattering of electrons from $\vec{e}d$ collisions.  In this way, they
will extract $G_M^s$, $G_E^s$, and $G_A^e$ separately, so that their
results for $G_M^s$ and $G_E^s$ will not be contaminated by uncertain
contributions to $G_A^e$.  These data will cover a range of $Q^2$ from
0.1 to 1.0 GeV$^2$, and thereby test the calculations of Zhu et.
al~\cite{Zhu:2000gn} over this range.

Another technique for avoiding the axial term is to observe the
parity-violating asymmetry in scattering from a spinless, isoscalar
target, like $^4$He, as proposed by Musolf and
Donnelly~\cite{Musolf:1992xm}.  In this case, only the electric form
factors contribute to the asymmetry.  Two measurements at Jefferson
Lab will make use of this idea, at $Q^2$ = 0.1 and 0.6~GeV$^2$, to
measure $G_E^s$.  The low $Q^2$ experiment~\cite{E00114} will measure
the slope of $G_E^s$, while the experiment at moderate
$Q^2$~\cite{E91004} will measure the actual value of $G_E^s$.

Elastic scattering of neutrinos from protons does not suffer the
difficulties described above, as there is no $G_A^{\rm ewm}$ form
factor in neutrino scattering.  This letter explores what can be
learned about the strange axial form factor by combining existing $\nu
p \rightarrow \nu p$ and $\bar{\nu} p \rightarrow \bar{\nu} p$ data
with the existing and upcoming data on parity-violating $\vec{e}p$ scattering.

The HAPPEX Collaboration has measured the forward-angle
parity-violating asymmetry in $\vec{e}p$ elastic
scattering at $Q^2 = 0.477$~GeV$^2$.  They report a combination of
$G_E^s$ and $G_M^s$~\cite{Aniol:2000at}:
$$G_E^s + 0.392G_M^s = 0.025 \pm 0.020 \pm 0.014.$$
Their result contains a contribution as well from $G_A^e$, but their
sensitivity to $G_A^e$ is $\sim$ 4\% of that to $G_E^s$ and $G_M^s$
because their measurement is at a very forward angle. They used a
value of $G_A^e$ from Ref.~\cite{Zhu:2000gn}.  The calculation of
Ref.~\cite{Zhu:2000gn} is not yet confirmed at this value of $Q^2$,
but even a large error in it would only produce a small additional
uncertainty in the interpretation of the HAPPEX result.

\begin{table}[t]
\caption{\label{E734_table} Differential cross section data from BNL
  E734~\protect\cite{Ahrens:1987xe}.  The uncertainties shown are
  total; they include statistical, $Q^2$-dependent systematic, and
  $Q^2$-independent systematic contributions, all added in quadrature.
  The extra row at the bottom (``0.50 GeV$^2$'') lists the cross sections averaged
  between the 0.45 and 0.55 GeV$^2$ points.}
\begin{ruledtabular}
\begin{tabular}{c|c|c}
$Q^2$ & $d\sigma/dQ^2(\nu p)$ & $d\sigma/dQ^2(\bar{\nu} p)$ \\
GeV$^2$ &  $10^{-12}$~(fm/GeV)$^2$ & $10^{-12}$~(fm/GeV)$^2$ \\
\hline
0.45 & $0.165  \pm0.033$  & $0.0756 \pm0.0164$ \\
0.55 & $0.109  \pm0.017$  & $0.0426 \pm0.0062$ \\
0.65 & $0.0803 \pm0.0120$ & $0.0283 \pm0.0037$ \\
0.75 & $0.0657 \pm0.0098$ & $0.0184 \pm0.0027$ \\
0.85 & $0.0447 \pm0.0092$ & $0.0129 \pm0.0022$ \\
0.95 & $0.0294 \pm0.0074$ & $0.0108 \pm0.0022$ \\
1.05 & $0.0205 \pm0.0062$ & $0.0101 \pm0.0027$ \\
\hline                            
0.50 & $0.137  \pm0.023$  & $0.0591 \pm0.0102$
\end{tabular}
\end{ruledtabular}
\end{table}

The only major measurement to date of $\nu p$ and $\bar{\nu} p$
elastic scattering cross sections took place at Brookhaven National
Laboratory, Experiment E734~\cite{Ahrens:1987xe}, using wide-band
neutrino and anti-neutrino beams of average kinetic energy 1.25 GeV
incident upon a large liquid scintillator target-detector system.
Table~\ref{E734_table} summarizes the results of this experiment.
Several attempts have been made to extract the strange axial form
factor from these
data~\cite{Ahrens:1987xe,Garvey:1993cg,Alberico:1998qw}.  In all
cases, there was an assumption made that $G_A^s$ had a dipole
$Q^2$-dependence --- that assumption will not be made here.

\begin{figure}[t]
\begin{center}
\scalebox{.6}{
\includegraphics*[145,490][520,720]{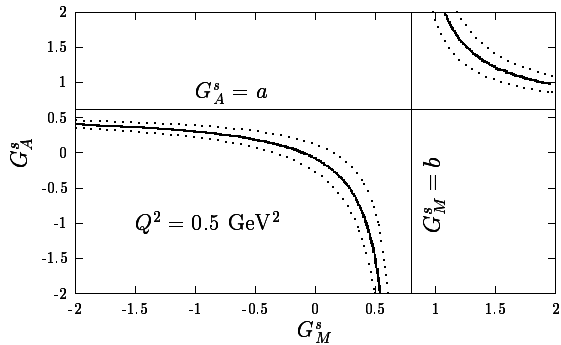}
}
\end{center}
\caption{Relation between $G_A^s$ and $G_M^s$ at $Q^2=0.5$
  (GeV)$^2$.  The solid line is the central value, while the dotted
  lines correspond to the total uncertainties in the E734 data.
  The asymptotes occurring for $G_A^s=a$ and $G_M^s=b$ are indicated,
  see text for details.}
\label{gms_gas_fig}
\end{figure}

Many years ago Llewellyn Smith~\cite{LlewellynSmith:1972zm} noted the
usefulness of measuring the difference in the differential cross
sections of the charged-current reactions
$\bar{\nu}n\rightarrow\mu^-p$ and $\nu p\rightarrow\mu^+n$, as this
yields a simple relation between $G_A^{CC}$ and the magnetic form
factors of the proton and neutron.  In the present discussion
$G_A^{CC}$ is regarded as known and the difference in the cross
sections of the neutral current processes $\nu p \rightarrow \nu p$
and $\bar{\nu} p \rightarrow \bar{\nu} p$ is used to relate $G_M^s$
and $G_A^s$.  The cross section for $\nu p$ and $\bar{\nu} p$ elastic
scattering is given by~\cite{Garvey:1993cg}
$$ \frac{d\sigma}{dQ^2} = \frac{G_F^2}{2\pi} \frac{Q^2}{E_\nu^2} (A\pm BW + CW^2)$$
where the $+$ ($-$) sign is for $\nu$ ($\bar{\nu}$) scattering, and
\begin{eqnarray*}
W    &=& 4(E_\nu /M_p - \tau) \\
\tau &=& Q^2/4M_p^2 \\
A    &=& \frac{1}{4}\left[(G_A^Z)^2(1+\tau)-\left((F_1^Z)^2-\tau(F_2^Z)^2\right)(1-\tau)\right.\\
     & & ~~~~~~~\left. +4\tau{F_1^Z}{F_2^Z}\right] \\
B    &=& -\frac{1}{4}G_A^Z(F_1^Z + F_2^Z) \\
C    &=& \frac{1}{64\tau}\left[(G_A^Z)^2 + (F_1^Z)^2 + \tau(F_2^Z)^2\right].
\end{eqnarray*}
Here, $E_\nu$ is the neutrino beam energy, and $F_1^Z$, $F_2^Z$, and
$G_A^Z$ are respectively the neutral weak Dirac, Pauli, and axial form
factors.  Taking the difference of these two cross sections,
$$
\Delta \equiv 
\frac{d\sigma}{dQ^2}(\nu p) - \frac{d\sigma}{dQ^2}(\bar{\nu} p) 
= - \frac{G_F^2}{4\pi} \frac{Q^2}{E_{\nu}^2} G_A^Z(F_1^Z + F_2^Z)W
$$
produces a relation between the Sachs magnetic form factors (in $F_1^Z
+ F_2^Z$) and the axial form factors (in $G_A^Z$).  By making use of
charge symmetry one may show that the sum of the weak neutral Dirac
and Pauli form factors is
$$F_1^Z + F_2^Z = \frac{1}{2}\left[(1-4\sin^2{\theta_W})G_M^p - G_M^n  - G_M^s\right].$$
Also recalling that
$$G_A^Z = \frac{1}{2}(-G_A^{CC} + G_A^s)$$
then the expression involving $\Delta$ may be written as:
\begin{eqnarray*}
\left[G_A^{CC}\right]G_M^s - G_A^sG_M^s 
+ \left[(1-4\sin^2{\theta_W})G_M^p - G_M^n\right]G_A^s \\
+ \left[\frac{16\pi}{W}\frac{E_\nu^2}{Q^2}\frac{\Delta}{G_F^2}
- G_A^{CC}[(1-4\sin^2{\theta_W})G_M^p - G_M^n]\right] = 0.
\end{eqnarray*}
This relationship has the form
$$ aG_M^s - G_A^sG_M^s + bG_A^s + c = 0 $$ 
where one may easily identify the factors $a$, $b$, and $c$ from the
previous equation.  Using the dipole form for the magnetic form
factors, $G_M^{p,n} = {\mu_{p,n}}/{(1+Q^2/M_V^2)^2}$ (where $M_V =
0.843$ GeV is the vector mass and $\mu_p = 2.793$ and $\mu_n = -1.913$
are the proton and neutron magnetic moments), and using the cross
sections from E734, one may calculate a relation between $G_M^s$ and
$G_A^s$, shown in Figure~\ref{gms_gas_fig} for $Q^2=0.5$~GeV$^2$.
There is an asymptote in $G_M^s$ when $G_A^s=a$, and similarly an
asymptote in $G_A^s$ when $G_M^s = b$.  Since the absolute values of
$G_M^s$ and $G_A^s$ are unlikely to be very large, this relation rules
out a range of moderate positive values of both $G_M^s$ and $G_A^s$.

\begin{figure}[t]
\begin{center}
\scalebox{.6}{
\includegraphics*[145,490][520,720]{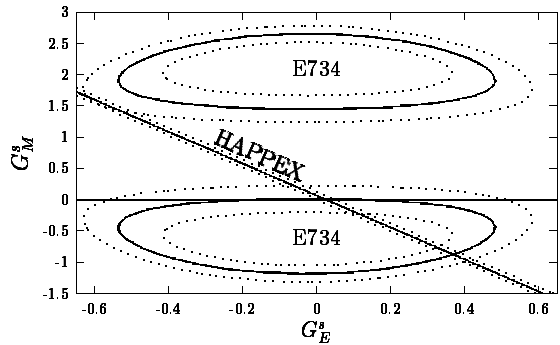}
}
\end{center}
\caption{The solid contours show the relation between $G_M^s$ and $G_E^s$ at
  $Q^2=0.5$~GeV$^2$, using the E734 data.  The dotted contours correspond to the
  total E734 uncertainties.  The straight (solid and dotted) lines
  show the HAPPEX result at $Q^2=0.477$~GeV$^2$.}
\label{gesgms_fig}
\end{figure}

If one now {\em adds} the $\nu$ and $\bar{\nu}$ cross sections
together, the dependence on $G_A^Z$ may be eliminated using the
expression derived from the difference of the cross sections, leaving
a relation between $F_1^Z$ and $F_2^Z$, and hence a relation between
$G_E^s$ and $G_M^s$:
\begin{eqnarray*}
\Sigma & \equiv & \frac{d\sigma}{dQ^2}(\nu p) + \frac{d\sigma}{dQ^2}(\bar{\nu} p)\\
       &=& \frac{G_F^2}{4\pi} \frac{Q^2}{E_{\nu}^2}\left[ 
         \left(-1+\tau+\frac{W^2}{16\tau}\right){\left(F_1^Z\right)}^2\right. \\
       & & +\left(+1+\tau+\frac{W^2}{16\tau}\right)\frac{(4\pi)^2\Delta^2}{G_F^4}\frac{E_{\nu}^4}{Q^4}\frac{1}{W^2(F_1^Z + F_2^Z)^2}\\
       & & +\left.\left(+1-\tau+\frac{W^2}{16\tau}\right)\tau{\left(F_2^Z\right)}^2 + 4\tau F_1^Z F_2^Z  \right].
\end{eqnarray*}
This relation can be expressed as a fourth-order polynomial in $G_E^s$
and $G_M^s$.  The solutions to this expression are contours in the
$(G_E^s,G_M^s)$ plane.  A set of contours, using the E734 data at
$Q^2=0.5$~GeV$^2$, is shown in Figure~\ref{gesgms_fig}.  The dipole
and Galster~\cite{Galster:1971kv} forms are used for the electric form
factors of the proton and neutron, respectively:
$$
G_E^p = \frac{1}{(1+Q^2/M_V^2)^2}, ~~~~~ G_E^n = -\frac{\mu_n \tau}{1+5.6\tau}G_E^p.
$$

\begin{figure}[thb]
\begin{center}
\scalebox{.6}{
\includegraphics*[145,490][520,720]{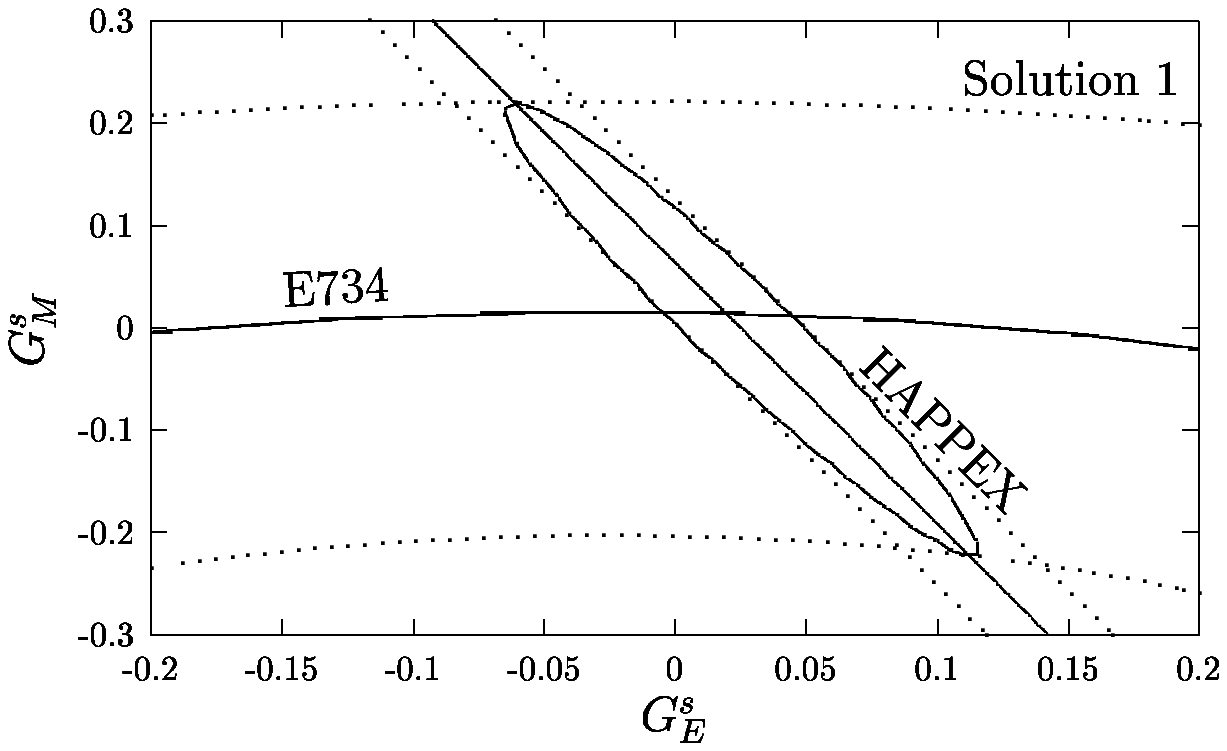}
}
\end{center}
\begin{center}
\scalebox{.6}{
\includegraphics*[145,490][520,720]{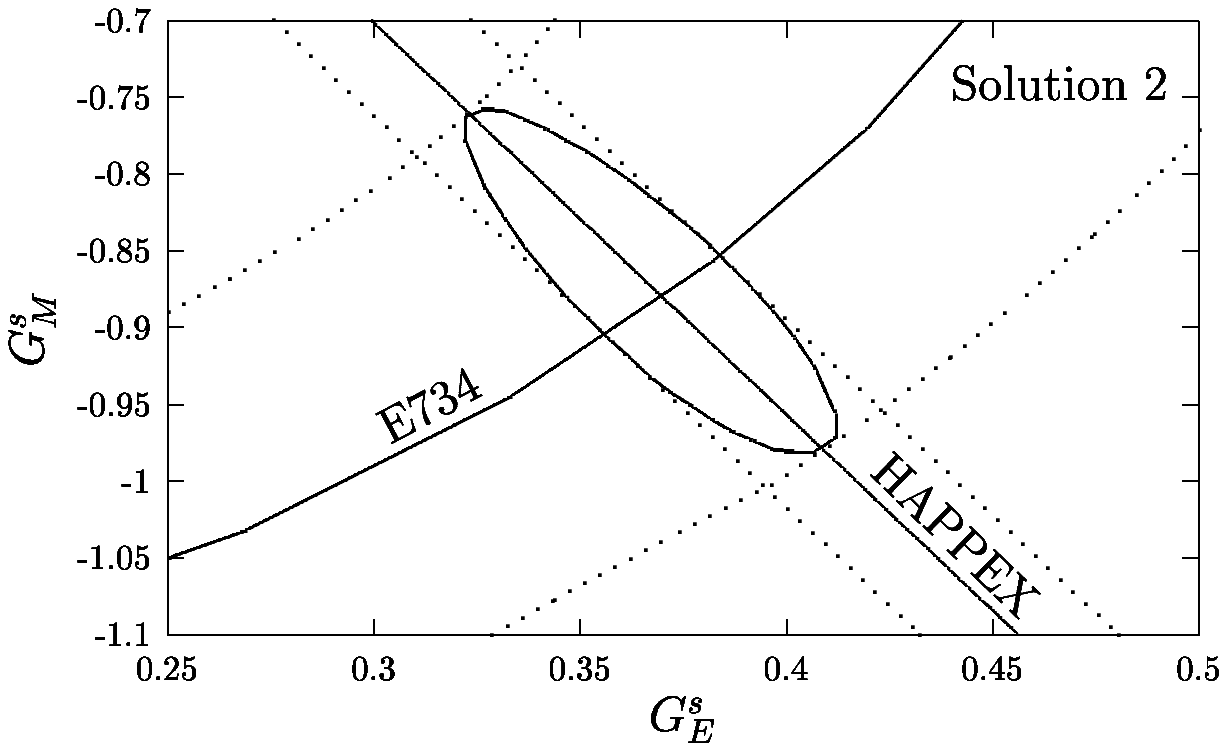}
}
\end{center}
\caption{Two solutions for $G_E^s$ and $G_M^s$ from the HAPPEX
  and E734 data.  The dotted lines correspond
  to the total errors of each experiment.  The solid contours
  bound the allowed regions of $G_E^s$ and $G_M^s$.}
\label{happex_e734_fig}
\end{figure}

These two relationships, one between $G_M^s$ and $G_A^s$, and another
between $G_M^s$ and $G_E^s$, need some additional input from another
experiment before any actual values of $G_E^s$, $G_M^s$, and $G_A^s$
can be determined.  The only existing additional experimental
information in this $Q^2$ range is the HAPPEX
result~\cite{Aniol:2000at}.  As displayed in Figs.~\ref{gesgms_fig}
and \ref{happex_e734_fig}, combining the HAPPEX results at
$Q^2=0.477$~GeV$^2$ with the E734 data at $Q^2=0.5$~GeV$^2$ gives two
solutions for $G_E^s$, $G_M^s$, and $G_A^s$, listed in
Table~\ref{solutions_table}.  This is the first determination of
$G_E^s$ and $G_A^s$ for $Q^2\not=0$.  These two solutions are very
different from each other, and in a few years one of them will be
selected with a measurement at a nearby $Q^2$, as will be done in the
$G^0$ and E91-004 experiments at Jefferson Lab.  However, there are
already good reasons to favor Solution~1 over Solution~2.  First of
all, the value of $G_A^s$ in Solution~1 is consistent with the
estimated value of $G_A^s(Q^2=0) = \Delta s = -0.14\pm 0.03$ from
deep-inelastic data~\cite{Filippone:2001ux}, whereas that found in
Solution~2 is much larger and of a different sign.  Similarly, the
value of $G_M^s$ in Solution~1 is consistent with that measured by
SAMPLE, whereas the value in Solution~2 is much larger in magnitude.
However, the final determination must come from the additional data to
be collected at Jefferson Lab in the next few years.

\begin{table}[t]
\caption{\label{solutions_table} Two solutions for the strange form
  factors at $Q^2=0.5$~GeV$^2$ produced from the E734 and HAPPEX data.}
\begin{tabular}{c|c|c}
      & Solution 1 & Solution 2 \\
\hline
$G_E^s$ & $~~0.02\pm 0.09$ & $~~0.37\pm 0.04$ \\ 
$G_M^s$ & $~~0.00\pm 0.21$ & $ -0.87\pm 0.11$ \\
$G_A^s$ & $ -0.09\pm 0.05$ & $~~0.28\pm 0.10$
\end{tabular}
\end{table}

The extensive program of measurements to be done by the $G^0$
Collaboration, when combined with the BNL E734 data, will give the
$Q^2$-dependence of $G_A^s$ in the range $0.45$--$0.95$~GeV$^2$.  The
first phase of the $G^0$ experiment (in 2005) will measure forward
scattering of protons (very similar to the HAPPEX measurement at
$Q^2=0.477$~GeV$^2$, and similarly insensitive to $G_A^e$) for $Q^2$
in the range $0.1$--$1.0$~GeV$^2$.  That will yield two solutions (in
a similar fashion as Figs.~\ref{gesgms_fig} and \ref{happex_e734_fig})
for each of $G_E^s$, $G_M^s$, and $G_A^s$ for several $Q^2$ points in
the range $0.45$--$0.95$~GeV$^2$.  The second phase of $G^0$ (in
2005-2006) will observe backward scattered electrons, and those data
will select the set of solutions to use.  The uncertainty in the
extraction of $G_A^s$ from the E734 and $G^0$ data will be between
$\pm0.03$ and $\pm0.10$, depending on the value of $G_M^s$ --- these
two are highly correlated, as Fig.~\ref{gms_gas_fig} shows.  These
data on $G_A^s$ will be crucial in nailing down the value of $\Delta
s$, which has been one of the goals of hadronic spin physics for many
years.

The author acknowledges M. Burkardt, G.T. Garvey, W.R. Gibbs,
and W.C. Louis for useful discussions; M.D. Marx for providing a copy
of E. Stern's thesis; and M. Diwan for providing E734
simulation code.  This work was supported by the US Department of Energy.

\bibliography{paper}

\begin{thebibliography}{22}
\expandafter\ifx\csname natexlab\endcsname\relax\def\natexlab#1{#1}\fi
\expandafter\ifx\csname bibnamefont\endcsname\relax
  \def\bibnamefont#1{#1}\fi
\expandafter\ifx\csname bibfnamefont\endcsname\relax
  \def\bibfnamefont#1{#1}\fi
\expandafter\ifx\csname citenamefont\endcsname\relax
  \def\citenamefont#1{#1}\fi
\expandafter\ifx\csname url\endcsname\relax
  \def\url#1{\texttt{#1}}\fi
\expandafter\ifx\csname urlprefix\endcsname\relax\def\urlprefix{URL }\fi
\providecommand{\bibinfo}[2]{#2}
\providecommand{\eprint}[2][]{\url{#2}}

\bibitem[{\citenamefont{Kaplan and Manohar}(1988)}]{Kaplan:1988ku}
\bibinfo{author}{\bibfnamefont{D.~B.} \bibnamefont{Kaplan}} \bibnamefont{and}
  \bibinfo{author}{\bibfnamefont{A.}~\bibnamefont{Manohar}},
  \bibinfo{journal}{Nucl. Phys.} \textbf{\bibinfo{volume}{B310}},
  \bibinfo{pages}{527} (\bibinfo{year}{1988}).

\bibitem[{\citenamefont{McKeown}(1989)}]{Mckeown:1989ir}
\bibinfo{author}{\bibfnamefont{R.~D.} \bibnamefont{McKeown}},
  \bibinfo{journal}{Phys. Lett.} \textbf{\bibinfo{volume}{B219}},
  \bibinfo{pages}{140} (\bibinfo{year}{1989}).

\bibitem[{\citenamefont{Beck}(1989)}]{Beck:1989tg}
\bibinfo{author}{\bibfnamefont{D.~H.} \bibnamefont{Beck}},
  \bibinfo{journal}{Phys. Rev.} \textbf{\bibinfo{volume}{D39}},
  \bibinfo{pages}{3248} (\bibinfo{year}{1989}).

\bibitem[{\citenamefont{Van~de Wiele and Morlet}(2003)}]{PVA4}
\bibinfo{author}{\bibfnamefont{J.}~\bibnamefont{Van~de Wiele}}
  \bibnamefont{and} \bibinfo{author}{\bibfnamefont{M.}~\bibnamefont{Morlet}}
  (\bibinfo{collaboration}{A4}), \bibinfo{journal}{Czech. Jour. Phys.}
  \textbf{\bibinfo{volume}{53}}, \bibinfo{pages}{A1} (\bibinfo{year}{2003}).

\bibitem[{G0:(2001)}]{G0:Beck}
\emph{\bibinfo{title}{The $G^0$ Experiment Backward Angle Measurements}},
  \bibinfo{organization}{The $G^0$ Collaboration, D. Beck, Spokesperson}
  (\bibinfo{year}{2001}).

\bibitem[{E00(2000)}]{E00114}
\bibinfo{organization}{Jefferson Lab Experiment E00-114, D. Armstrong and R.
  Michaels, Spokespersons} (\bibinfo{year}{2000}).

\bibitem[{E91(1991)}]{E91004}
\bibinfo{organization}{Jefferson Lab Experiment E91-004, E. Beise,
  Spokesperson} (\bibinfo{year}{1991}).

\bibitem[{E99(1999)}]{E99115}
\bibinfo{organization}{Jefferson Lab Experiment E99-115, K. Kumar and D.
  Lhuillier, Spokespersons} (\bibinfo{year}{1999}).

\bibitem[{\citenamefont{Musolf et~al.}(1994)}]{Musolf:1994tb}
\bibinfo{author}{\bibfnamefont{M.~J.} \bibnamefont{Musolf}}
  \bibnamefont{et~al.}, \bibinfo{journal}{Phys. Rept.}
  \textbf{\bibinfo{volume}{239}}, \bibinfo{pages}{1} (\bibinfo{year}{1994}).

\bibitem[{\citenamefont{Liesenfeld et~al.}(1999)}]{Liesenfeld:1999mv}
\bibinfo{author}{\bibfnamefont{A.}~\bibnamefont{Liesenfeld}}
  \bibnamefont{et~al.} (\bibinfo{collaboration}{A1}), \bibinfo{journal}{Phys.
  Lett.} \textbf{\bibinfo{volume}{B468}}, \bibinfo{pages}{20}
  (\bibinfo{year}{1999}).

\bibitem[{\citenamefont{Filippone and Ji}(2001)}]{Filippone:2001ux}
\bibinfo{author}{\bibfnamefont{B.~W.} \bibnamefont{Filippone}}
  \bibnamefont{and} \bibinfo{author}{\bibfnamefont{X.-D.} \bibnamefont{Ji}},
  \bibinfo{journal}{Advances in Nuclear Physics} \textbf{\bibinfo{volume}{26}},
  \bibinfo{pages}{1} (\bibinfo{year}{2001}).

\bibitem[{\citenamefont{Musolf and Holstein}(1990)}]{Musolf:1990ts}
\bibinfo{author}{\bibfnamefont{M.~J.} \bibnamefont{Musolf}} \bibnamefont{and}
  \bibinfo{author}{\bibfnamefont{B.~R.} \bibnamefont{Holstein}},
  \bibinfo{journal}{Phys. Lett.} \textbf{\bibinfo{volume}{B242}},
  \bibinfo{pages}{461} (\bibinfo{year}{1990}).

\bibitem[{\citenamefont{Musolf and Donnelly}(1992)}]{Musolf:1992xm}
\bibinfo{author}{\bibfnamefont{M.~J.} \bibnamefont{Musolf}} \bibnamefont{and}
  \bibinfo{author}{\bibfnamefont{T.~W.} \bibnamefont{Donnelly}},
  \bibinfo{journal}{Nucl. Phys.} \textbf{\bibinfo{volume}{A546}},
  \bibinfo{pages}{509} (\bibinfo{year}{1992}).

\bibitem[{\citenamefont{Zhu et~al.}(2000)\citenamefont{Zhu, Puglia, Holstein,
  and Ramsey-Musolf}}]{Zhu:2000gn}
\bibinfo{author}{\bibfnamefont{S.-L.} \bibnamefont{Zhu}},
  \bibinfo{author}{\bibfnamefont{S.~J.} \bibnamefont{Puglia}},
  \bibinfo{author}{\bibfnamefont{B.~R.} \bibnamefont{Holstein}},
  \bibnamefont{and} \bibinfo{author}{\bibfnamefont{M.~J.}
  \bibnamefont{Ramsey-Musolf}}, \bibinfo{journal}{Phys. Rev.}
  \textbf{\bibinfo{volume}{D62}}, \bibinfo{pages}{033008}
  (\bibinfo{year}{2000}).

\bibitem[{\citenamefont{Hasty et~al.}(2000)}]{Hasty:2001ep}
\bibinfo{author}{\bibfnamefont{R.}~\bibnamefont{Hasty}} \bibnamefont{et~al.}
  (\bibinfo{collaboration}{SAMPLE}), \bibinfo{journal}{Science}
  \textbf{\bibinfo{volume}{290}}, \bibinfo{pages}{2117} (\bibinfo{year}{2000}).

\bibitem[{\citenamefont{Ito}(2003)}]{Ito:2003mr}
\bibinfo{author}{\bibfnamefont{T.~M.} \bibnamefont{Ito}}
  (\bibinfo{collaboration}{SAMPLE}) (\bibinfo{year}{2003}),
  \eprint{nucl-ex/0310001}.

\bibitem[{\citenamefont{Ahrens et~al.}(1987)}]{Ahrens:1987xe}
\bibinfo{author}{\bibfnamefont{L.~A.} \bibnamefont{Ahrens}}
  \bibnamefont{et~al.}, \bibinfo{journal}{Phys. Rev.}
  \textbf{\bibinfo{volume}{D35}}, \bibinfo{pages}{785} (\bibinfo{year}{1987}).

\bibitem[{\citenamefont{Aniol et~al.}(2001)}]{Aniol:2000at}
\bibinfo{author}{\bibfnamefont{K.~A.} \bibnamefont{Aniol}} \bibnamefont{et~al.}
  (\bibinfo{collaboration}{HAPPEX}), \bibinfo{journal}{Phys. Lett.}
  \textbf{\bibinfo{volume}{B509}}, \bibinfo{pages}{211} (\bibinfo{year}{2001}).

\bibitem[{\citenamefont{Garvey et~al.}(1993)\citenamefont{Garvey, Louis, and
  White}}]{Garvey:1993cg}
\bibinfo{author}{\bibfnamefont{G.~T.} \bibnamefont{Garvey}},
  \bibinfo{author}{\bibfnamefont{W.~C.} \bibnamefont{Louis}}, \bibnamefont{and}
  \bibinfo{author}{\bibfnamefont{D.~H.} \bibnamefont{White}},
  \bibinfo{journal}{Phys. Rev.} \textbf{\bibinfo{volume}{C48}},
  \bibinfo{pages}{761} (\bibinfo{year}{1993}).

\bibitem[{\citenamefont{Alberico et~al.}(1999)}]{Alberico:1998qw}
\bibinfo{author}{\bibfnamefont{W.~M.} \bibnamefont{Alberico}}
  \bibnamefont{et~al.}, \bibinfo{journal}{Nucl. Phys.}
  \textbf{\bibinfo{volume}{A651}}, \bibinfo{pages}{277} (\bibinfo{year}{1999}).

\bibitem[{\citenamefont{Llewellyn~Smith}(1972)}]{LlewellynSmith:1972zm}
\bibinfo{author}{\bibfnamefont{C.~H.} \bibnamefont{Llewellyn~Smith}},
  \bibinfo{journal}{Phys. Rept.} \textbf{\bibinfo{volume}{3}},
  \bibinfo{pages}{261} (\bibinfo{year}{1972}).

\bibitem[{\citenamefont{Galster et~al.}(1971)}]{Galster:1971kv}
\bibinfo{author}{\bibfnamefont{S.}~\bibnamefont{Galster}} \bibnamefont{et~al.},
  \bibinfo{journal}{Nucl. Phys.} \textbf{\bibinfo{volume}{B32}},
  \bibinfo{pages}{221} (\bibinfo{year}{1971}).

\end{thebibliography}

\end{document}